%% file: Main.tex
\documentclass[journal ]{new-aiaa}
\usepackage[utf8]{inputenc}
\usepackage{textcomp}

\usepackage{graphicx}
\usepackage{amsmath}
\usepackage[version=4]{mhchem}
\usepackage{siunitx}
\usepackage{longtable,tabularx}
\title{Effect of span size in Scale Resolving Simulations of airfoil stall and post-stall}

\author{Francesco M. D'Afiero \footnote{Master student, Department of Industrial Engineering.} and Renato Tognaccini\footnote{Professor, Department of Industrial Engineering.}}
\affil{University of Naples Federico II, Naples, Italy}
\author{Gianluca Iaccarino\footnote{Professor, Department of Mechanical Engineering.}}
\affil{University of Stanford, USA}

\input{Configuration/config}

\begin{document}

\maketitle

\input{Sections/Body}

\input{Sections/Appendix/Appendix}

\end{document}

%% file: Configuration/config.tex
\usepackage[utf8]{inputenc}
\usepackage{graphicx}
\usepackage{amsmath}
\usepackage[version=4]{mhchem}
\usepackage{siunitx}
\usepackage{silence}
\usepackage[export]{adjustbox}
\usepackage{graphicx} 
\usepackage[section]{placeins}
\usepackage{multirow}
\usepackage{etoolbox}
\usepackage{float}
\usepackage{printlen}
\usepackage{subdepth}
\usepackage{todonotes}
\usepackage[labelformat=simple]{subcaption}
\usepackage{tcolorbox}
\usepackage{booktabs}
\usepackage[normalem]{ulem}
\useunder{\uline}{\ul}{}
\usepackage[section]{placeins}

\MakeRobust{\subref}

\newsavebox{\leftimage}
\newsavebox{\middleimage}
\newsavebox{\rightimage}
\newlength{\imageheight}

\usepackage{dcolumn}%
\usepackage{bm}%

\mathcode`\/="202F
\usepackage{tikz}
\usepackage{scalerel}
\usepackage{xcolor}

\newcommand{\tikzcircle}[1][black]{\scalerel*{\tikz \draw[line width=0.5pt, color=#1, fill=white] (0,0) circle (2pt);}{\circ}}

\definecolor{mplcolorA}{HTML}{1f77b4}
\definecolor{mplcolorB}{HTML}{ff7f0e}
\definecolor{mplcolorC}{HTML}{2ca02c}
\definecolor{mplcolorD}{HTML}{d62728}
\definecolor{mplcolorE}{HTML}{9467bd}
\definecolor{mplcolorF}{HTML}{8c564b}
\definecolor{mplcolorG}{HTML}{e377c2}
\definecolor{mplcolorH}{HTML}{7f7f7f}
\definecolor{mplcolorI}{HTML}{bcbd22}
\definecolor{mplcolorJ}{HTML}{17becf}
\definecolor{mplcolorK}{HTML}{17becf}

\newcommand{\pltmarkerA}[1][mplcolorA,fill=mplcolorA]{%
  \scalerel*{%
    \begin{tikzpicture}
      \fill[#1] (0,0) rectangle (4pt, 4pt);
    \end{tikzpicture}%
  }{\tikzcircle}%
}
\newcommand{\pltmarkerB}[1][mplcolorB,fill=mplcolorB]{%
  \scalerel*{%
    \begin{tikzpicture}
      \fill[#1] (0,0) -- (1.5pt, 2.5pt) -- (3pt,0) -- cycle;
    \end{tikzpicture}%
  }{\tikzcircle}%
}
\newcommand{\pltmarkerC}[1][mplcolorC,fill=mplcolorC]{%
  \scalerel*{%
    \begin{tikzpicture}
      \fill[#1] (0,0) -- (1.5pt, -2.5pt) -- (3pt,0) -- cycle;
    \end{tikzpicture}%
  }{\tikzcircle}%
}
\newcommand{\pltmarkerD}[1][mplcolorD,fill=mplcolorD]{%
  \scalerel*{%
    \begin{tikzpicture}
      \draw[rounded corners=0.1pt,#1,fill] (0,-2.5pt)--++(0,5pt)--++(-180+30:5pt)--cycle;
    \end{tikzpicture}%
  }{\tikzcircle}%
}
\newcommand{\pltmarkerE}[1][mplcolorE,fill=mplcolorE]{%
  \scalerel*{%
    \begin{tikzpicture}
      \filldraw[rounded corners=0.1pt,#1] (0,2.5pt)--++(0,-5pt)--++(-180+30:-5pt)--cycle;
    \end{tikzpicture}%
  }{\tikzcircle}%
}
\newcommand{\pltmarkerF}[1][mplcolorF,fill=mplcolorF]{%
  \scalerel*{%
    \begin{tikzpicture}
      \fill[#1] (0,0) rectangle (4pt, 4pt);
    \end{tikzpicture}%
  }{\tikzcircle}%
}

%% file: Sections/Body.tex
The use of computational fluid dynamics (CFD) tools in aerodynamic design has a long history. Approaches based on Reynolds-Averaged Navier-Stokes (RANS) and Unsteady RANS (URANS)
turbulence models have considerable limitations in the presence of massively separated flow regions,, and this has lead to
more and more emphasis on "scale-resolving" methods; in this work we focus on Large Eddy Simulations (LES) that are becoming more affordable thanks to advances in numerical methods and the increased availability of computer resources.
Computational
Fluid Dynamics (CFD) practices involve Reynolds-Averaged Navier-Stokes (RANS) and Unsteady RANS (URANS)
turbulence models that are ill-equipped to accurately predict massively separated flows;
LES are therefore becoming an extremely valuable tool for design purposes.

A key question pertains to large scale separated flows  if they can be effectively represented statistically in a two-dimensional manner, the last observation being important to assess to appropriately define the aerodynamic behaviour of wing sections at any angle of attack. In the case of stalled airfoils, even experiments are extremely challenging because of the presence of significant large-scale structures which likely interact with the spanwise boundary of wind tunnels, thereby introducing genuine 3D effects. When conducting large scale simulations, the influence of the physical confinement are absent as long as the spanwise size is sufficiently large. In such cases, the primary concern revolves around the adequacy of the spanwise resolution and the averaging process. It is important to note that if a large 3D structure exists, capturing a single instance of it could be insufficient. Multiple structures need to be captured to ensure a statistically valid 2D representation.

When it comes to numerical methods for addressing such challenging regimes, scale resolving simulations and specifically  Implicit Large Eddy Simulation (ILES) can be particularly suitable. However, the demanding computational effort required for very accurate analyses of airfoil flows, force the researchers to limit the spanwise size of the adopted grid to few percents of the airfoil chord (here we notice that the computational cost scales linearly with the spanwise extent because uniform meshes are typically used in the span).
 The study presented here gains significance considering the lack of a clear trend or justification in the literature for the chosen span extent beyond the necessity of limiting computational resources. Table \ref{LiteratureSimulations} summarizes the adopted spanwise extent $l_z$ for LES of various airfoil flows at different chord Reynolds number, $Re_c$, and angles of attack, $\alpha$.

The effects on the difference of the flow physics resolved by different spanwise extent of the computational domain were addressed for the cylinder case at a very low Reynolds number in \cite{garcia2019span} where it was pointed out that the magnitude of the vortex-stretching term has a primary effect on the generation of the three-dimensional motion. Park et. al \cite{park2017high} observed a significant effect of spanwise extent in the case of airfoil simulation at a single angle of attack ($\alpha = 60^\circ$) which is in the region of the very deep stall. No systematic analysis of the span size on the scale resolving simulations of airfoil flows is reported in the literature  when trying to correctly predict the more critical stall and early post-stall phenomenology: this is the aim of present note.

\begin{table}[htbp]
\caption{Summary of scale-resolving airfoil simulations with the adopted spanwise dimension.}
\label{LiteratureSimulations}
\begin{center}
\begin{tabular}{llll}
\hline
 & $Re_c$ & $\alpha \left[deg\right]$ & $ l_z$ \\ \hline
Streher \cite{streher2018} & $1x10^5$ & 5, 11 & $0.25c$ \\
Park J.S. et al. \cite{park2017high} & $2.7x10^5$ & 60 & $[c - 7c]$ \\
Sato et al \cite{sato2019mechanisms}. & $1.6x10^6$ & 20.1 & $0.05c$ \\
Schmidt et al. \cite{schmidt20} & $1.64x10^6$ & 12 & $0.05c$ \\
Park G. et al. \cite{park2014improved}& $1.64x10^6$ & 12 & $0.72c$ \\
Frère et al. \cite{frere2018high} & $1.64x10^6$ & 12 & $0.01c$ \\
Dahlstrom et al. \cite{dahlstrom2001large}& $2.1x10^6$ & 13.3 & $0.12c$ \\
Kawai et al. \cite{kawai2013wall} & $2.1x10^6$ & 13.3 & $0.017c$ \\
Mary et al. \cite{mary2002large}& $2.1x10^6$ & 13.3 & $[0.005c - 0.03c]$ \\
\hline
\end{tabular}
\end{center}
\label{Literature Simulations}
\end{table}

The airfoil analysed in this study is the NREL S826, commonly found in wind turbine applications, operating at a chord Reynolds number of $Re=10^5$. Experimental testing of this airfoil was conducted by Chivae et al. \citep{chivaee2014large}. In the same paper the authors also provided  numerical data obtained by Ellipsys3D, a LES solver with a dynamic subgrid model. Experimental measurements of the aerodynamic forces are available in both corrected and uncorrected forms. %
The correction addresses both solid blockage, which reduces the effective area of the test section due to the presence of the test model, and wake blockage, which is proportional to the wake size.

The computational framework employed in this study is based on the PyFR solver \cite{witherden2014pyfr}.
The simulations conducted are referred to as Implicit Large Eddy Simulations (ILES), where numerical dissipation solely models the subgrid-scale terms. Turbulence is not artificially introduced at the inlet or in the vicinity of the airfoil.
Fourth-order polynomials are utilized, resulting in a fifth-order accurate spatial solution. Time integration is carried out using a five-stage Runge-Kutta scheme of fourth-order accuracy with adaptive global time-stepping, selecting a baseline Courant number of 0.8.
Previous studies \cite{park2017high} demonstrated PyFR accuracy in critical conditions such as deep stall.

\begin{figure}[htbp]
    \centering
    \includegraphics[width=0.5\textwidth]{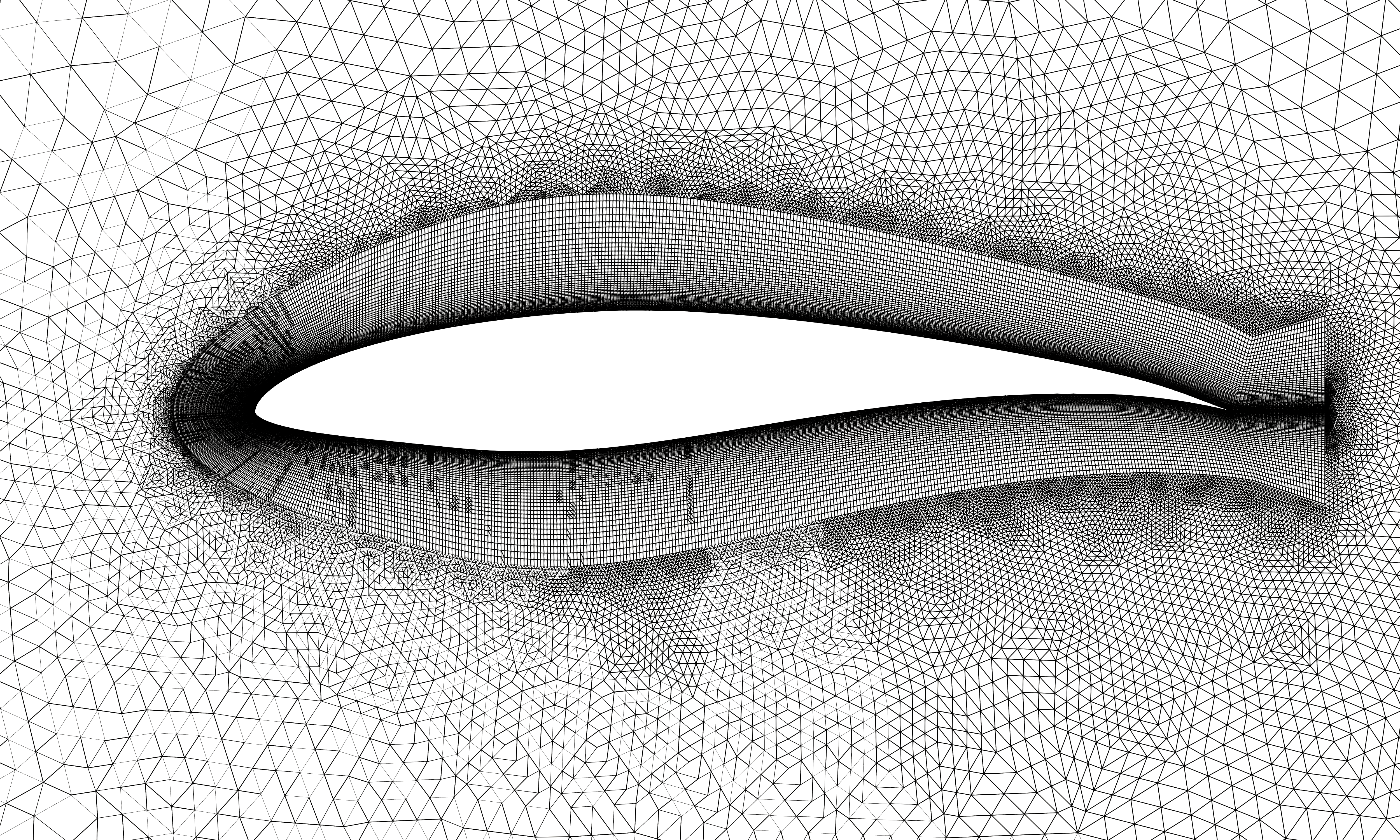}
    \caption{Numerical grid used for this study. The various cases only differ for the spanwise extent and resolution.}
    \label{fig:ILES Grids}
\end{figure}

All ILES computations presented herein are performed on a 2D mixed-element mesh extruded in the spanwise direction, with varying spanwise extents discussed for each case. The flow asymptotic conditions are set such that the freestream Mach number is 0.15 (compressible solver). The geometry and the 2D grid are generated using Gmsh \cite{gmsh}, featuring a C-type mesh with quadrangular elements near the airfoil and triangular elements in the farfield. Grids are structured near the airfoil and gradually coarsened where lower gradients are expected, reducing computational cost while maintaining a farfield extension of 20 chords. A two-dimensional view of one of the  employed grid is provided in Figure \ref{fig:ILES Grids}. Grid resolution is designed based on the requirements outlined by Choi and Moin \cite{choi2012grid}.\\

In this study we focus on evaluating statistically averaged aerodynamic forces; the sufficiency of the time-averaging in the simulations is assessed using the Augmented Dickey-Fuller (ADF) test \cite{said1984testing} applied to lift ($C_L$) and drag ($C_D$) coefficients.
Oscillations around the time-averaged $C_L$ and $C_D$ were treated as random noise without trend.
Calculations were conducted using CINECA Marconi100 GPU based HPC platform and Shepard and Yellowstone GPU clusters at Stanford University. Clearly the computational time strongly depends on the adopted grid size. The largest simulation consists of about $60 \cdot 10^6$ grid points solved with $60$ NVIDIA V100 GPUs leading to a wall time of roughly $48$ hours.

\begin{figure}[htbp]
    \centering
    \includegraphics[width=0.5\textwidth]{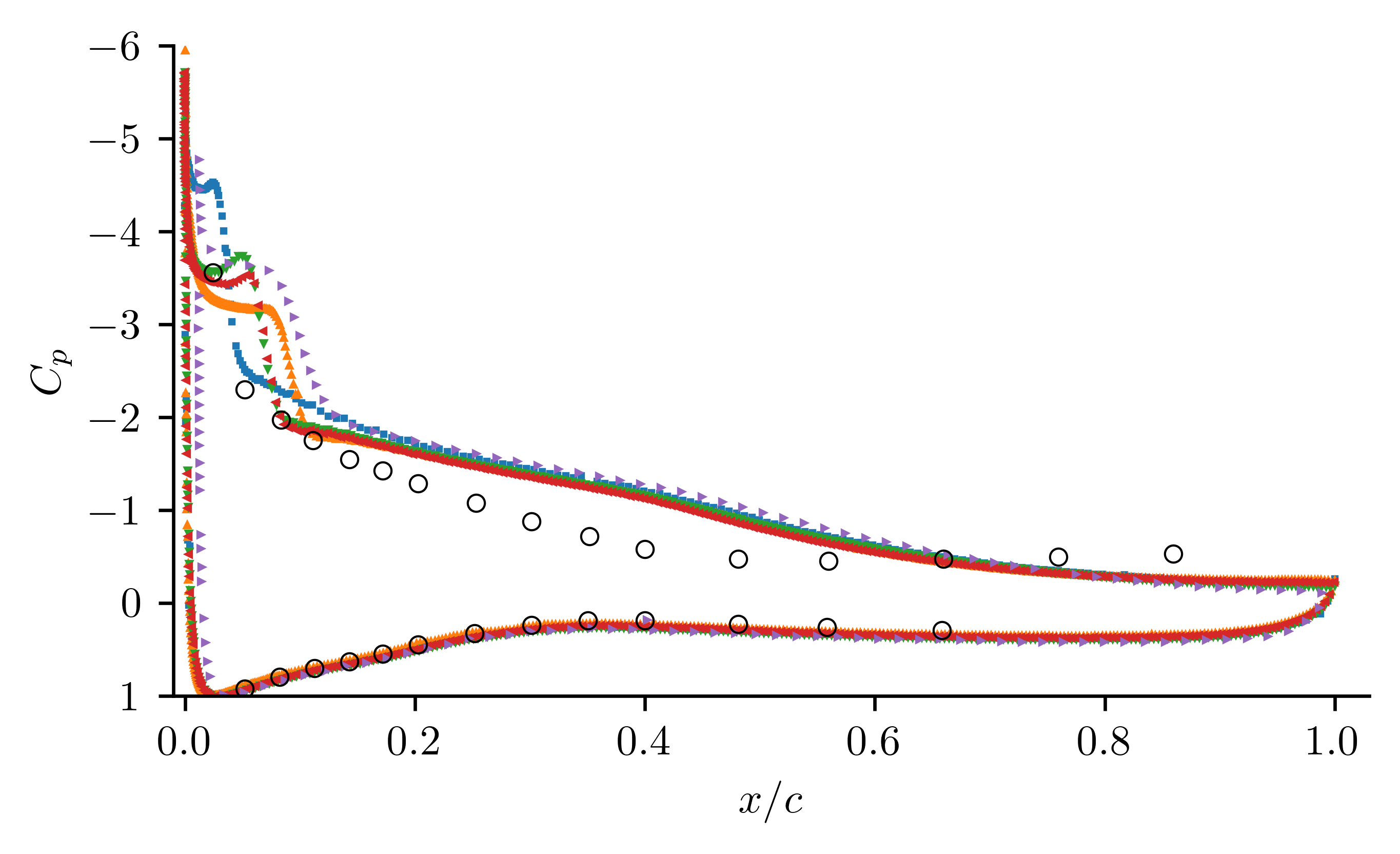}
    \caption{$C_p$ vs $x/c$ for NREL S826 with $Re_c = 10^5$, $M=0.15$ at $\alpha=12^\circ$: \protect\pltmarkerA \, ILES A, \protect\pltmarkerB \, ILES B, \, \protect\pltmarkerC \, ILES C, \protect\pltmarkerD \, ILES D, \protect\pltmarkerE \, EllipSys3D, \protect\tikzcircle \, Experiments. Ellipsys3D data are reported as in \cite{frere2015cross} Fig. 11.}
    \label{fig:Cp12degcomparison}
\end{figure}

The first set of simulations corresponds to $\alpha = 12^\circ$; simulations were performed using $l_z = 0.12c$ and $l_z = 0.72c$.
Table \ref{tb:LESsetups12deg} summarizes the results in terms of $C_L$ and $C_D$ as well as the characteristic element lengths of the grids. To facilitate comparison with classical refinement criteria of the finite volume and finite difference frameworks, each physical element length is divided by the polynomial order used for the ILES. In the same table the experimental and computational reference data are also reported. In this case, nearly at stall angle, the computed lift and drag coefficients weakly depend on the employed span size and are in  substantial agreement with the reference experimental and ILES data. Figure \ref{fig:Cp12degcomparison} shows the spanwise averaged surface pressure coefficient for the different ILES and the reference experimental and numerical data. Present ILES and reference calculations all show the presence of a laminar separation bubble with a negligible effect of the adopted span extent. The experiment does not show the laminar separation bubble, whereas indicates the presence of larger separated region in the aft part of the airfoil. It should however considered that in the experiment the pressure was just measured in a single section and not averaged along the span. Incidentally the present ILES and the reference Ellypsis3D simulations are in better agreement with the uncorrected experimental  lift force.

\begin{table}[htbp]
\centering
\caption{Summary of ILES setups and Lift and Drag coefficients comparisons for NREL S826 with $Re_c = 10^5$, $M=0.15$ at $\alpha=12^\circ$.}
\begin{tabular}{@{}ccccccccc@{}}
\hline
Setup & $l_z$ & \multicolumn{2}{c}{Element Length}    & $C_L$  & $\sigma_{CL}$ & $C_D$ & $\sigma_{CD}$\\
      &       & $\Delta y /c$   & $\Delta z /c$    &           &        \\
\hline
Exp.  Corrected    &       &          &                    & 1.1961 & & 0.1106 \\
Exp.  Not Corrected&       &          &                   & 1.4169 & & -      \\
\hline
Ellipsys3D & $1c$    &                   &                 & 1.5231 & & 0.0647 \\
\hline
ILES A     & $0.12c$ & 0.000616         & 0.0150           & 1.4843 & 0.0160 & 0.0625 & 0.0029\\
ILES B     & $0.12c$ & 0.000616      & 0.0075           & 1.4121 & 0.0147 & 0.0613 & 0.0027 \\
ILES C     & $0.72c$  & 0.000616         & 0.0450           & 1.4966 & 0.0093 & 0.0639 & 0.0030 \\
ILES D     & $0.72c$  & 0.000616         & 0.0225           & 1.4560 & 0.0065 & 0.0620 &0.0018 \\
\hline
\end{tabular}
\label{tb:LESsetups12deg}
\end{table}

\begin{table}[htbp]
\centering
\caption{Summary of ILES setups and Lift and Drag coefficients comparisons for NREL S826 with $Re_c = 10^5$, $M=0.15$ at $\alpha=20^\circ$.}
\begin{tabular}{@{}ccccccccc@{}}
\hline
Setup & $l_z$ & \multicolumn{2}{c}{Element Length} & $C_L$  & $\sigma_{CL}$ & $C_D$ & $\sigma_{CD}$  \\
      &       & $\Delta y /c$   & $\Delta z /c$    &         & &        & \\
\hline
Exp.  Corrected    &       &          &                   & 0.9894 & & 0.3757 & \\
Exp.  Not Corrected&       &          &                   & 1.1012 & & -     & \\
\hline
Ellipsys3D & $1c$    &                   &                & 1.2706 & & 0.4436 & \\
\hline
ILES A     & $0.06c$ & 0.000616         & 0.0075          & 1.7006 & 0.3794 & 0.6292 & 0.1825 \\
ILES B     & $0.12c$ & 0.000616         & 0.0075           & 1.5757 & 0.3076 & 0.6192 & 0.1381\\
ILES C     & $0.1395c$ & 0.000616      & 0.0465            & 1.7673 & 0.3555 & 0.6437 & 0.1800\\
ILES D     & $1c$  & 0.000616         & 0.0111           & 1.0107 & 0.0591 & 0.3745 & 0.0200 \\
ILES E     & $2c$  & 0.000616         & 0.0465            & 1.0777 & 0.0425 & 0.3978 &0.0133\\

\hline
\end{tabular}
\label{tb:LESsetups}
\end{table}

The second result discussed in detail is the $20^\circ$ angle of attack case. Following the experimental information, we are already in the post-stall region.
The various ILES setups, along with their corresponding results and comparisons with numerical and experimental data, are summarized in Table \ref{tb:LESsetups}.
The computed lift coefficient varies from $C_L = 1.7006$ for $l_z = 0.06c$ to $C_L = 1.0107$ at $l_z = 1c$. The dramatic influence of the spanwise size even leads to a non stalled airfoil for $l_z \le 0.1395c$, only small differences appear for the simulation with $l_z = 1c$ and $l_z = 2c$ making us confident that, at least for this case, $l_z = 1c$ is sufficient to get a converged lift independent of span. Lastly, also the drag coefficient shows large variations with the adopted $l_z$.

\begin{figure}[htbp]
    \centering
    \includegraphics[width=0.5\textwidth]{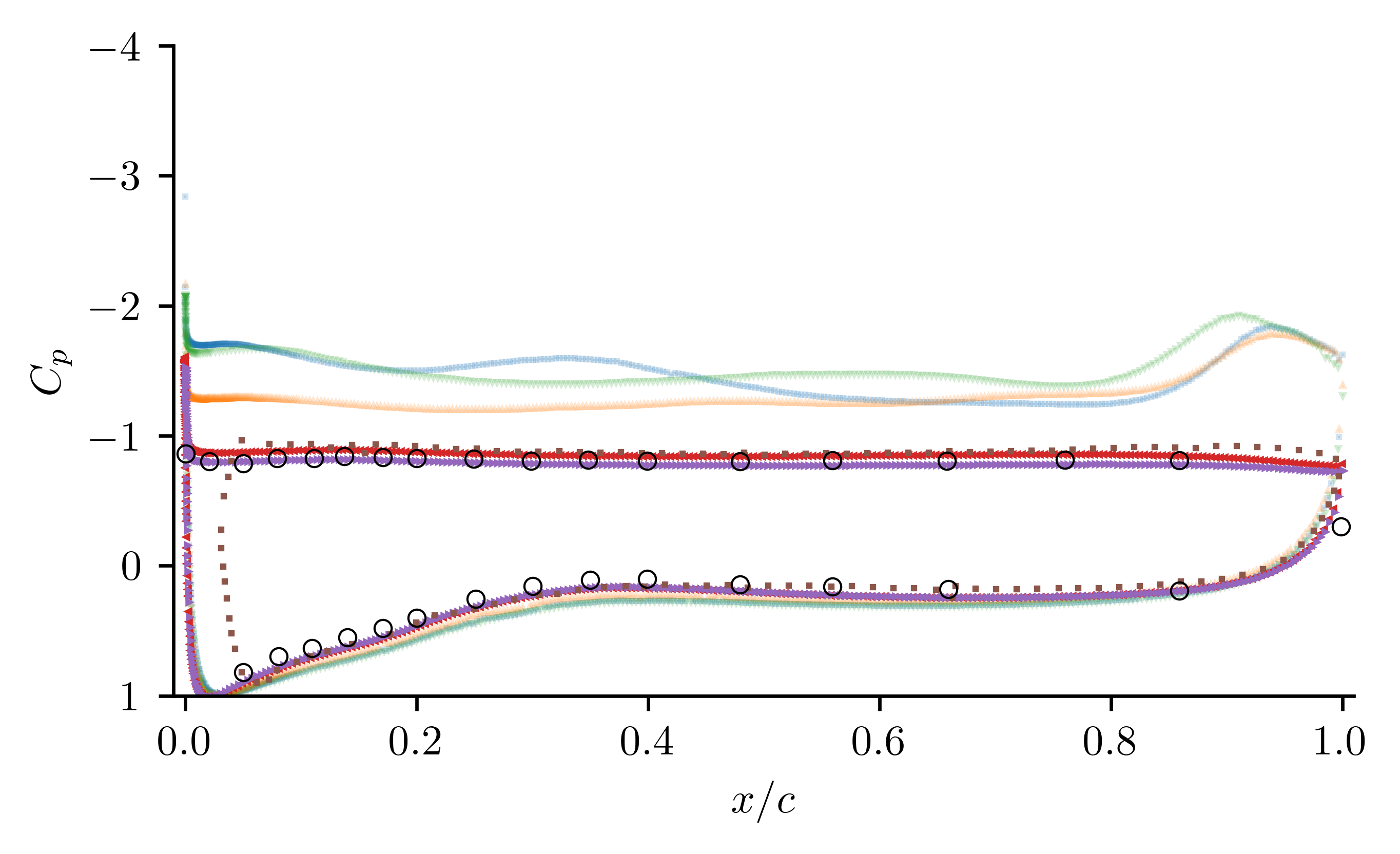}
    \caption{$C_p$ vs $x/c$ for NREL S826 with $Re_c = 10^5$, $M=0.15$ at $\alpha=20^\circ$: \protect\pltmarkerB \, ILES A, \protect\pltmarkerA \, ILES B, \, \protect\pltmarkerC \, ILES C, \protect\pltmarkerE \, ILES D, \protect\pltmarkerD \, ILES E, \protect\pltmarkerF \, EllipSys3D,  \protect\tikzcircle \, Experiments. Among the different ILES, only ILES D and ILES E are shown without reducing opacity. Ellipsys3D data are reported as in \cite{chivaee2014large} Fig. 4.20 (d).}
    \label{fig:Cp20deg}
\end{figure}

Figure \ref{fig:Cp20deg} reports the surface pressure coefficient distribution. It illustrates that the two larger span simulations with $l_z = 1c$ and $l_z = 2c$ exhibit consistent agreement with the experiment. Incidentally our large span ILES are in better agreement than the reference LES with the measured aerodynamic forces. All ILES simulations show a nearly flat pressure coefficient on the suction side of the airfoil making us confident that in all cases suction side flow is massively separated, however the simulations with $l_z <= 0.1395c$ are in offset with the reference experiments, this is the reason of the higher lift coefficient.

\begin{figure}[htbp]
     \centering

    \begin{subfigure}{0.32\textwidth}
    \includegraphics[width=\textwidth]{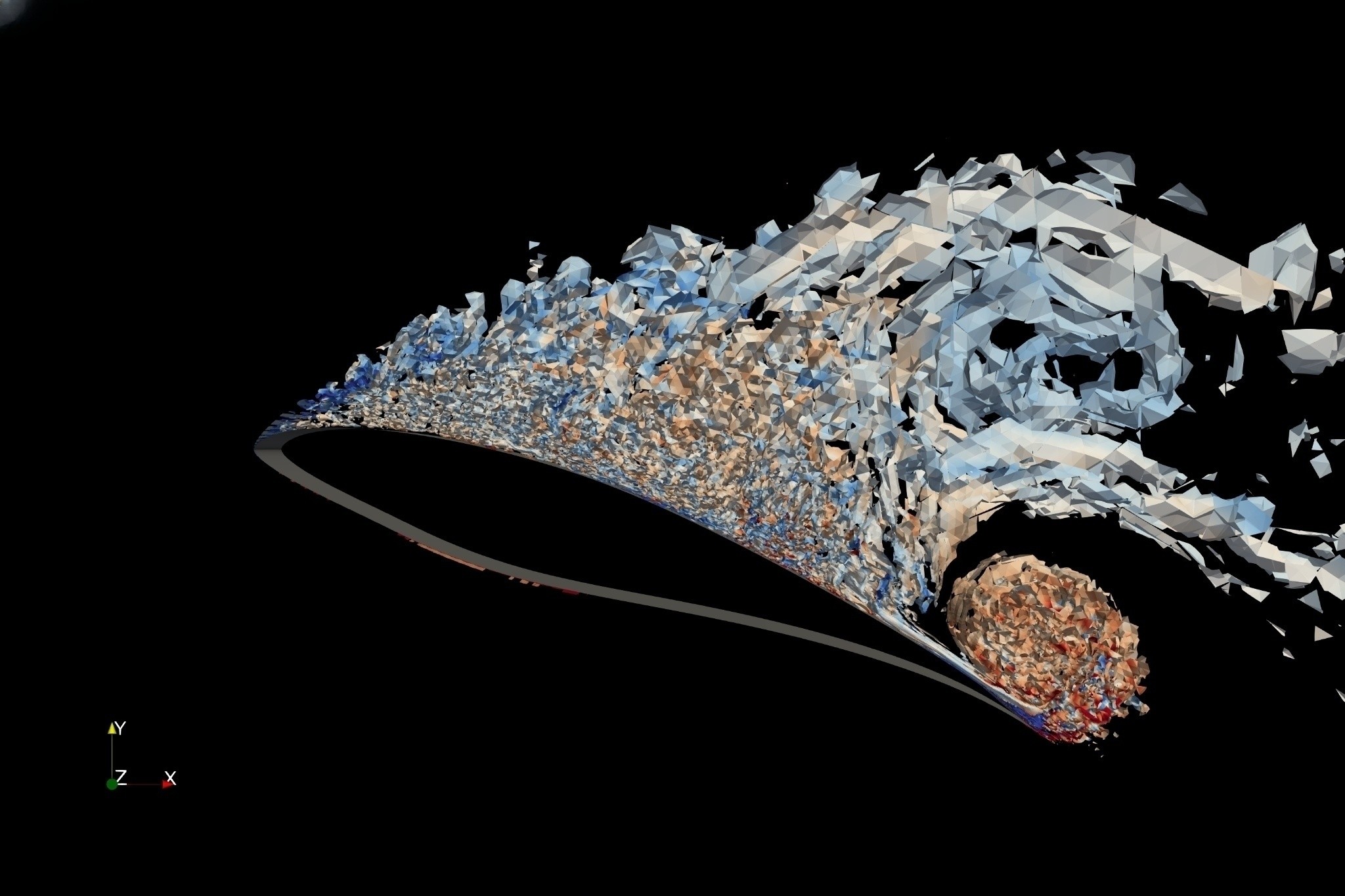}
    \caption{ }
    \label{fig:Qcrit_lz01395c}
    \end{subfigure}
    \begin{subfigure}{0.32\textwidth}
    \includegraphics[width=\textwidth]
    {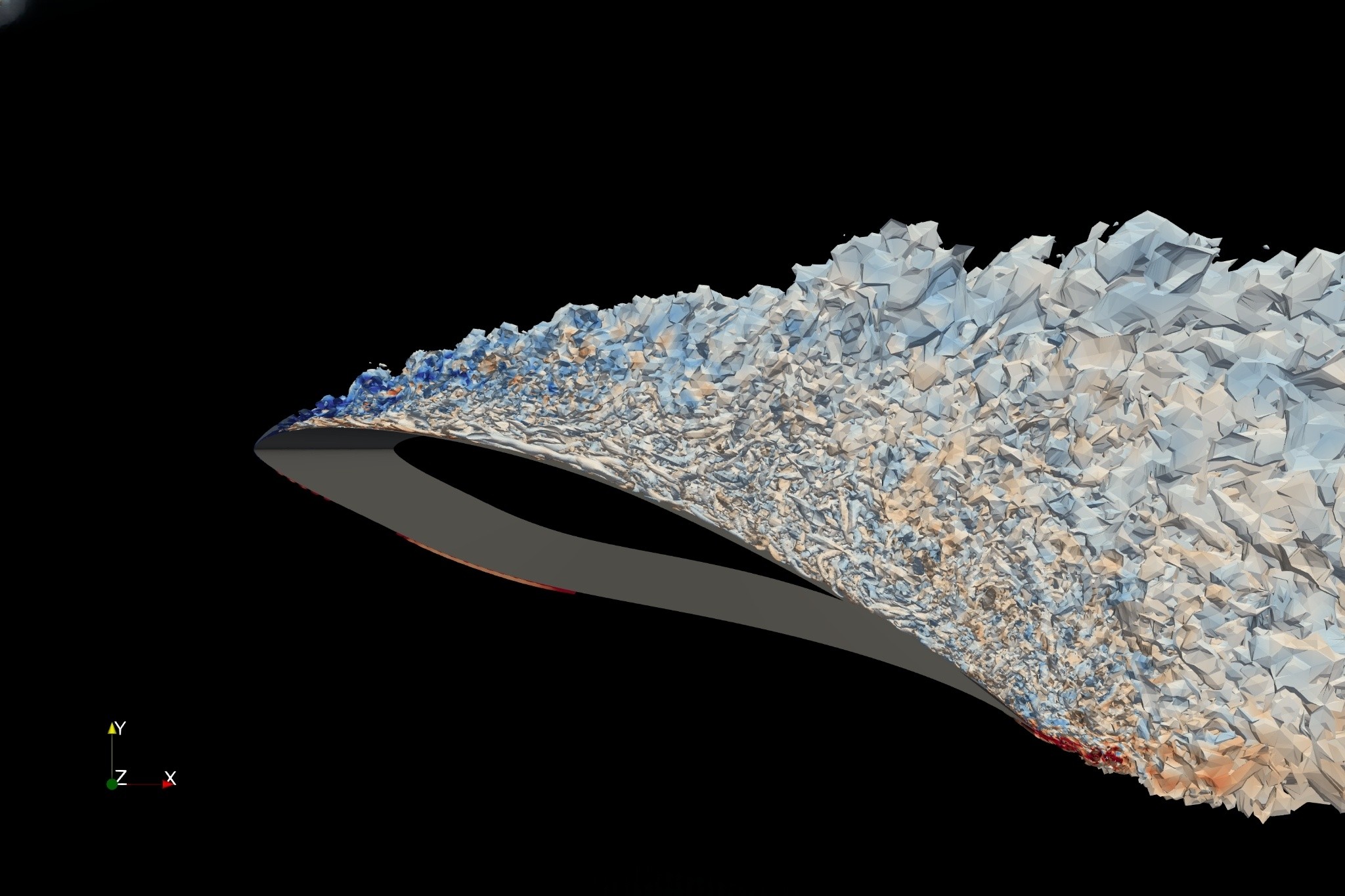}
    \caption{ }
    \label{fig:Qcrit_lz1c}
    \end{subfigure}
    \begin{subfigure}{0.32\textwidth}
    \includegraphics[width=\textwidth]{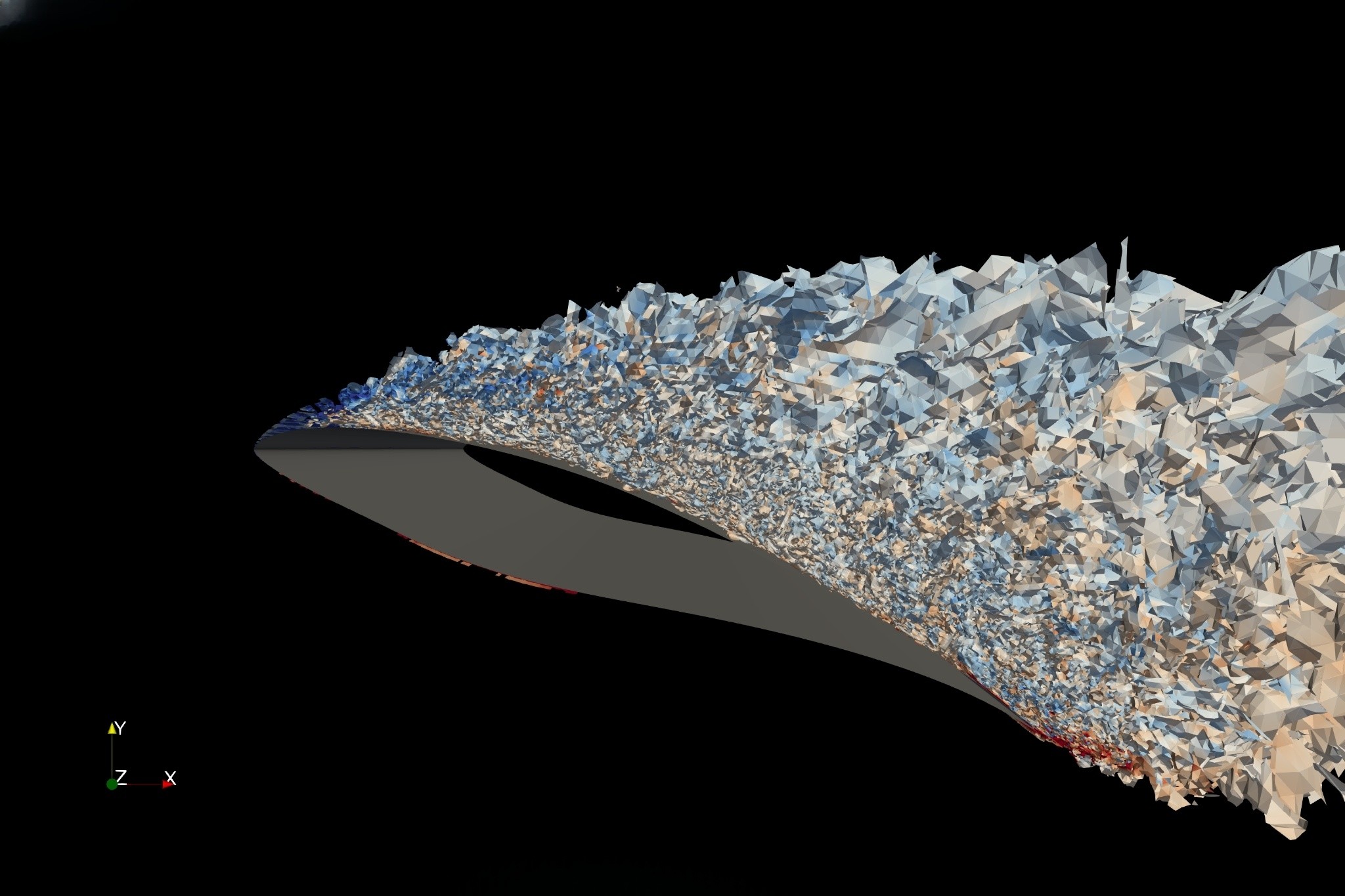}
    \caption{ }
    \label{fig:Qcrit_lz2c}
    \end{subfigure}

    \caption{Instantaneous isosurfaces of the second invariant of the velocity gradient tensor at time $t = 13 \,s$ colored by nondimensional spanwise vorticity $\omega_z c/ U_{\infty}$. Left column ILES C, middle column ILES B, right column ILES A.}
    \label{fig:Qcrit20deg}
\end{figure}

To understand the reason of this offset, it is interesting to analyze the isosurfaces of the second invariant of the velocity gradient tensor which are depicted in Figure \ref{fig:Qcrit20deg}(a), Figure \ref{fig:Qcrit20deg}(b) and Figure \ref{fig:Qcrit20deg} (c)  respectively for $l_z = 0.1395c$, $l_z = 1c$ and $l_z = 2c$. Figure \ref{fig:Qcrit_lz01395c} shows a persisting vortex on the suction side of the airfoil: it is the responsible of the lower pressure region on the airfoil which keeps high the computed lift coefficient even if the flow is massively separated. Figure \ref{fig:Qcrit_lz1c} and \ref{fig:Qcrit_lz2c} show that such vortex breaks down when larger span sizes are adopted, allowing for an higher pressure on the suction side of the airfoil and a much lower $C_L$ as consequence.

\begin{figure}[htbp]
    \centering
    \includegraphics[width=0.5\textwidth]{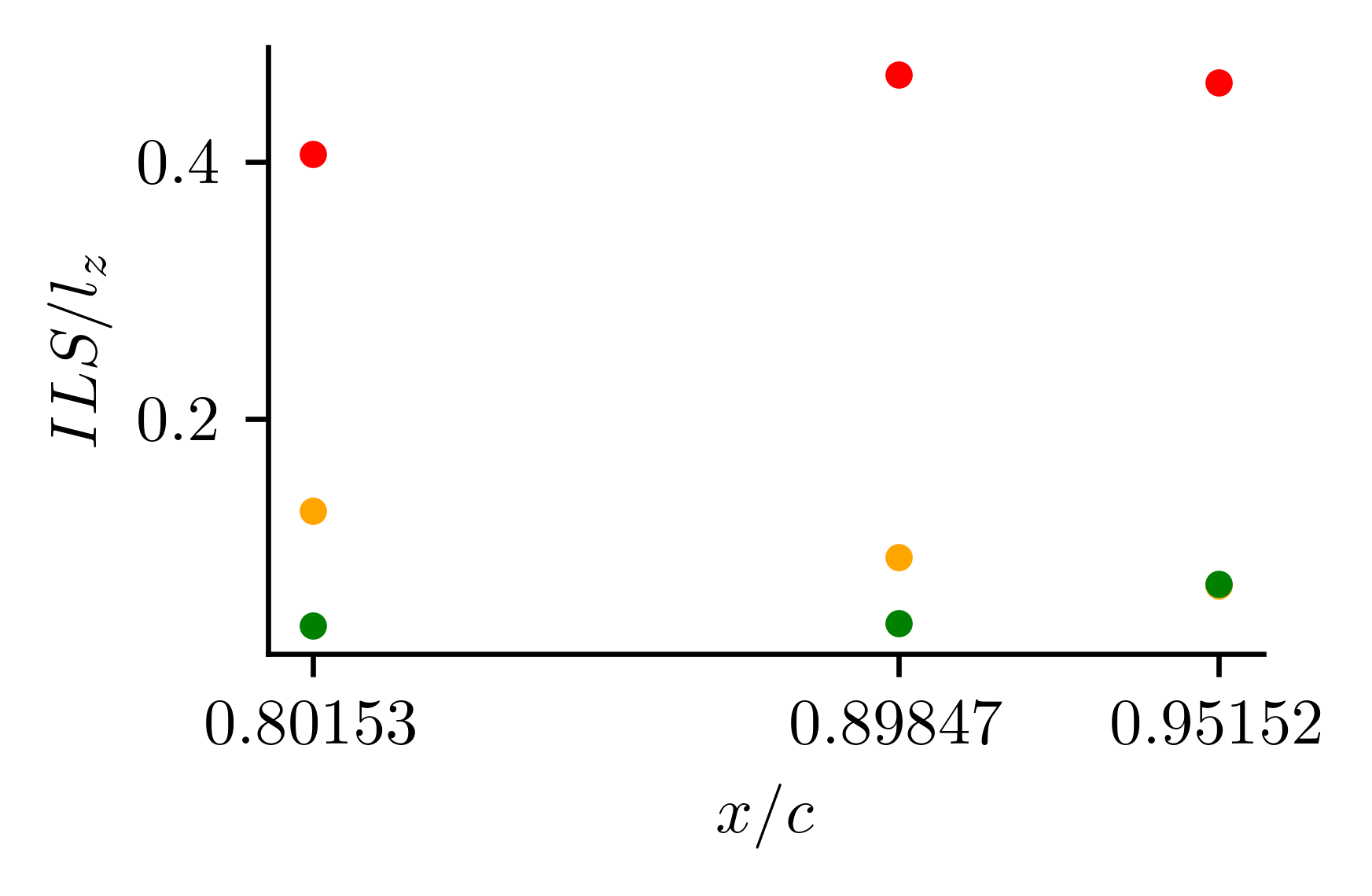}
    \caption{$ILS/l_z$ for the ILES with $l_z =0.1395c$ in red, ILES with $l_z=1c$ in orange and ILES with $l_z = 2c$ in green. At $x/c = 0.95152$ the ILES with $l_z=1c$ and $l_z = 2c$ are overlapping.}
    \label{fig:ILSvsX}
\end{figure}

This result is quantitatively confirmed via the analysis of the spanwise autocorrelations for the velocity fluctuations. Particularly we define a spanwise integral length scale as:

\begin{equation}
    ILS = \int_{0}^{z_{0.05}} R_{uu} \left( \xi \right) \, d\xi
\end{equation}

where $z_{0.05}$ is the $z$ at which $R_{uu}$ first crosses the $0.05$ threshold value here chosen to define the two signals as sufficiently uncorrelated. Figure \ref{fig:ILSvsX} depicts the values of $ILS/l_z$ at different $x/c$ locations for various span extents, respectively $l_z = 0.1395c$, $l_z = 1c$ and $l_z = 2c$,  at $\alpha = 20^{\circ}$. To compute the spanwise autocorrelations, we sample the velocity signals at a height within the boundary layer such that $y^+ = 30$ across all streamwise locations. This ensures that the correlated velocity signal falls within the buffer layer region. Such a metric enables us to observe that the effect of spanwise extent is such that it tends to approach zero with an increasing $l_z$ (theoretically infinite). However, if this metric shows significant changes between two vastly different spanwise extents, it indicates the insufficiency of the smaller span. Empirically, we find that a value of $ILS/l_z \approx 0.1$ serves as a suitable criterion for defining a sufficient spanwise size, corroborated by the aforementioned analysis.

The same study was repeated for the case $\alpha=25^\circ$ was: it is not reported here because confirms   the same results.

\begin{figure}[htbp]
    \begin{subfigure}{0.45\textwidth}
    \centering
    \includegraphics[width=\textwidth]{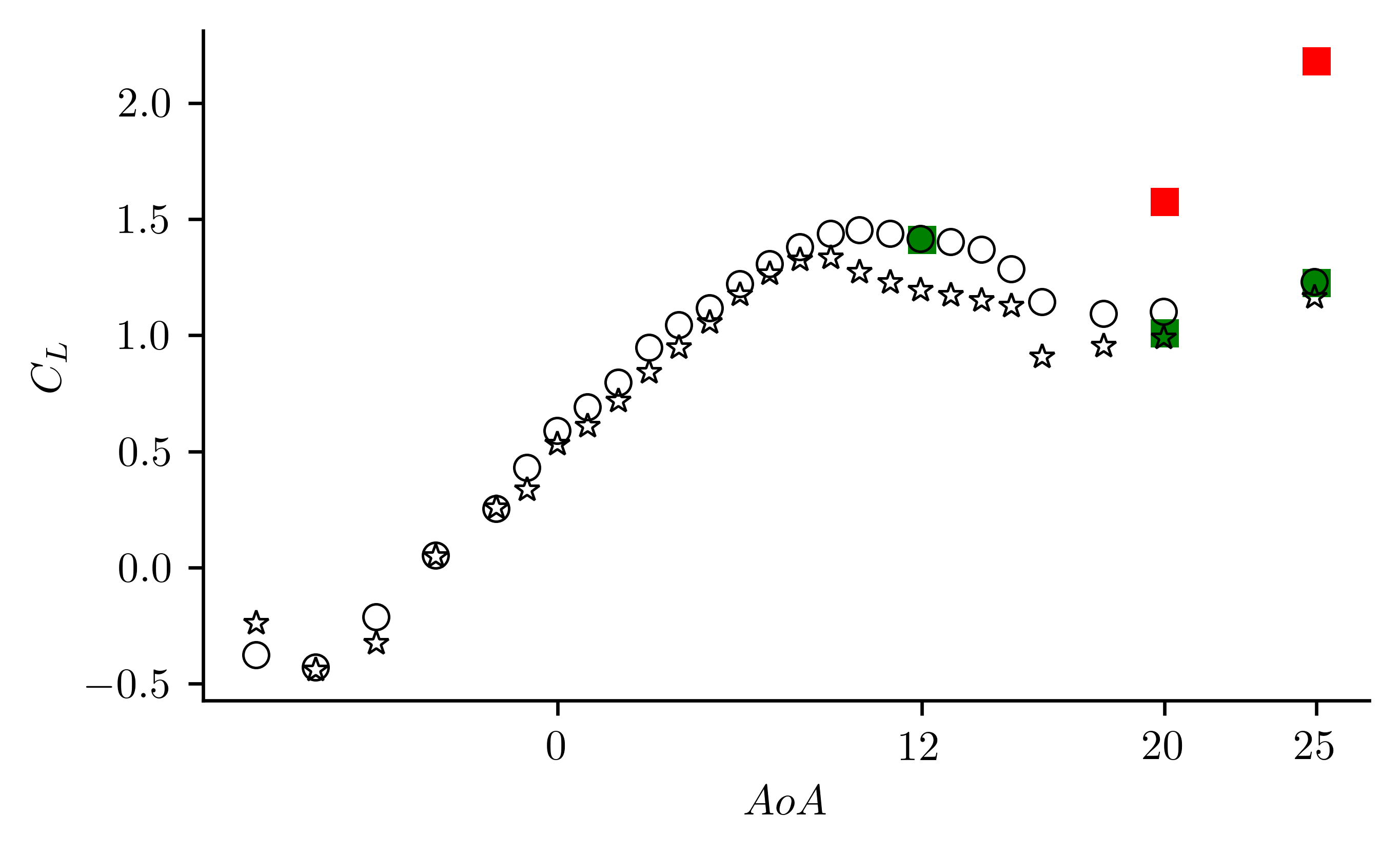}
    \end{subfigure}
    \begin{subfigure}{0.45\textwidth}
    \centering
    \includegraphics[width=\textwidth]{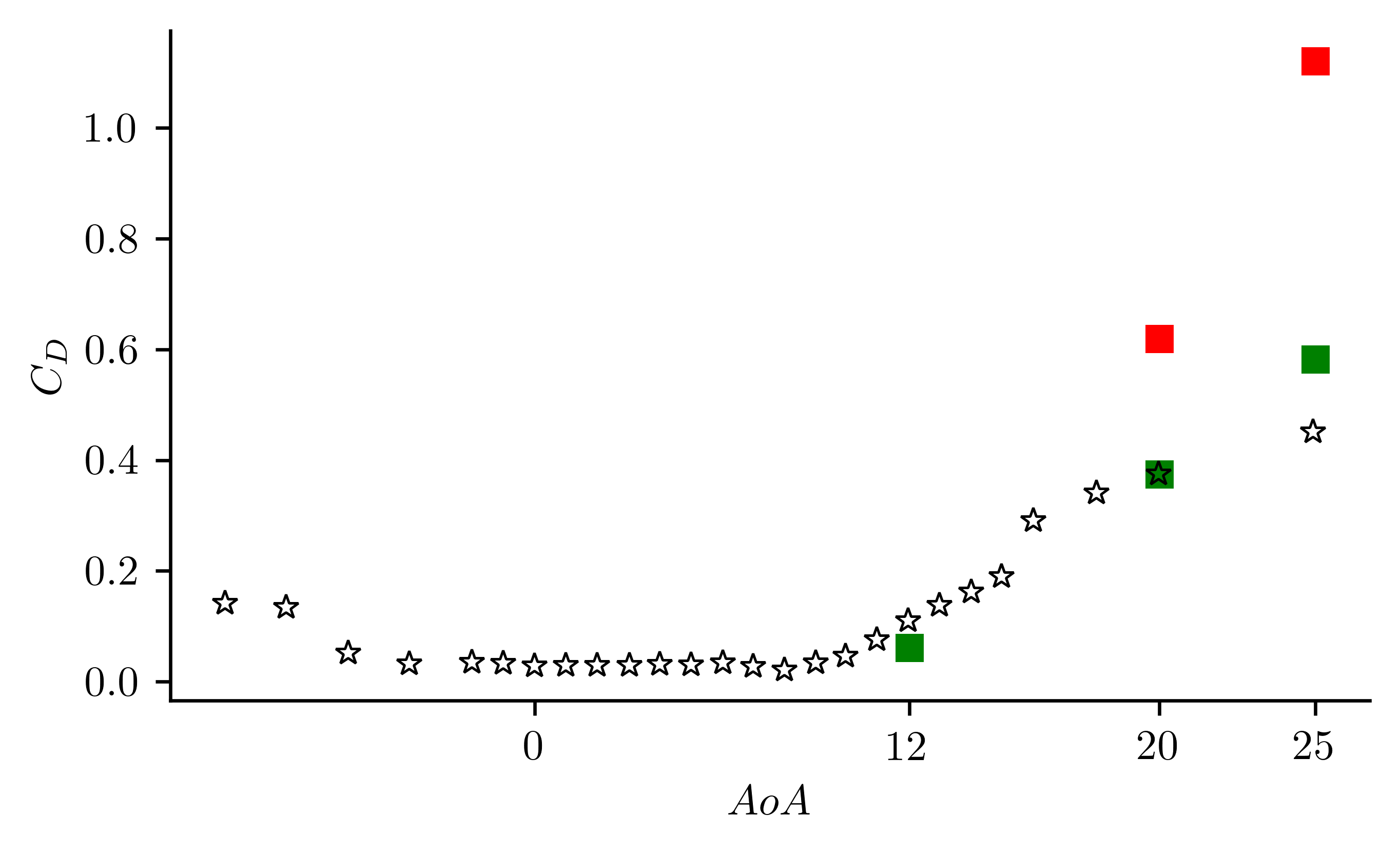}
    \end{subfigure}
    \caption{$C_L$ and $C_D$ vs $AoA$ for the NRELS826 at $M=0.2$ and $Re=10^5$. Circles denote non corrected measurements while stars denote the corrected ones. Red squares denote ILES with $l_z = 0.12c$, while green squares are $l_z = 1c$}
    \label{fig:CLCDvsExp}
\end{figure}

Figure \ref{fig:CLCDvsExp} shows the curves $C_L = C_L\left( \alpha \right)$ and $C_D = C_D\left( \alpha \right)$ comparing experiment with ILES $l_z=0.12c$ and $l_z=1c$.  While the numerical results overlap at $\alpha=12^\circ$, the effect of spanwise size is very clear for $\alpha=20^\circ$ and $\alpha=25^\circ$: the agreement with the experiment requires the adoption of $l_z=1c$ at least. In this case both $C_L$ and
$C_D$  well match measured data (interestingly again uncorrected values better than the corrected ones).

In conclusion, in this short note,  a parametric study was conducted on the influence of spanwise extent in ILES predictions of airfoils performances up to deep stall. For small angles of attack sufficient agreement with the experiment was observed with the usual 111recommendation of $l_z \approx 0.1c$. However, for massively separated flows, such as in stall and post-stall, a dramatic influence of the adopted spanwise extent was found. The analysis of the turbulent autocorrelations in the spanwise direction suggests that using a criterion derived from the integral length scale is more suitable than a purely geometrical one and a value  $l_z \approx 1c$  has established as a baseline for accurate simulations of  airfoil performances at high angle of attack by mean of ILES simulations.

\section*{Acknowlegments}

We acknowledge the CINECA award under the ISCRA initiative, for the availability of high-performance computing resources and support.

\FloatBarrier

%% file: Main.bbl
\begin{thebibliography}{16}
\newcommand{\enquote}[1]{``#1''}
\providecommand{\natexlab}[1]{#1}
\providecommand{\url}[1]{\texttt{#1}}
\providecommand{\urlprefix}{URL }
\expandafter\ifx\csname urlstyle\endcsname\relax
  \providecommand{\doi}[1]{\discretionary{}{}{}https://doi.org/#1}\else
  \providecommand{\doi}[1]{\discretionary{}{}{}\urlstyle{rm}\url{https://doi.org/#1}}\fi

\bibitem[{Garcia et~al.(2019)Garcia, Weymouth, Nguyen, and Tutty}]{garcia2019span}
Garcia, B.~F., Weymouth, G.~D., Nguyen, V.-T., and Tutty, O.~R., \enquote{Span effect on the turbulence nature of flow past a circular cylinder,} \emph{Journal of Fluid Mechanics}, Vol. 878, 2019, pp. 306--323.

\bibitem[{Park et~al.(2017)Park, Witherden, and Vincent}]{park2017high}
Park, J., Witherden, F., and Vincent, P., \enquote{High-order implicit large-eddy simulations of flow over a NACA0021 aerofoil,} \emph{AIAA Journal}, Vol.~55, No.~7, 2017, pp. 2186--2197.

\bibitem[{Streher(2018)}]{streher2018}
Streher, L.~B., \enquote{Large-eddy simulations of the flow around a NACA0012 airfoil at different angles of attack,} \emph{arXiv preprint arXiv:1807.01576}, 2018.

\bibitem[{Sato et~al.(2019)Sato, Asada, Nonomura, Aono, Yakeno, and Fujii}]{sato2019mechanisms}
Sato, M., Asada, K., Nonomura, T., Aono, H., Yakeno, A., and Fujii, K., \enquote{Mechanisms for turbulent separation control using plasma actuator at Reynolds number of 1.6$\times$ 106,} \emph{Physics of Fluids}, Vol.~31, No.~9, 2019, p. 095107.

\bibitem[{Schmidt et~al.(2001)Schmidt, Franke, and Thiele}]{schmidt20}
Schmidt, S., Franke, M., and Thiele, F., \enquote{Assessment of SGS models in LES applied to a NACA 4412 airfoil,} \emph{39th Aerospace Sciences Meeting and Exhibit}, 2001, p. 434.

\bibitem[{Park and Moin(2014)}]{park2014improved}
Park, G.~I., and Moin, P., \enquote{An improved dynamic non-equilibrium wall-model for large eddy simulation,} \emph{Physics of Fluids}, Vol.~26, No.~1, 2014, pp. 37--48.

\bibitem[{Fr{\`e}re et~al.(2018)Fr{\`e}re, Hillewaert, Chatelain, and Winckelmans}]{frere2018high}
Fr{\`e}re, A., Hillewaert, K., Chatelain, P., and Winckelmans, G., \enquote{High Reynolds number airfoil: from wall-resolved to wall- modeled LES,} \emph{Flow, Turbulence and Combustion}, Vol. 101, No.~2, 2018, pp. 457--476.

\bibitem[{Dahlstrom and Davidson(2001)}]{dahlstrom2001large}
Dahlstrom, S., and Davidson, L., \enquote{Large eddy simulation of the flow around an airfoil,} \emph{39th Aerospace Sciences Meeting and Exhibit}, 2001, p. 425.

\bibitem[{Kawai and Asada(2013)}]{kawai2013wall}
Kawai, S., and Asada, K., \enquote{Wall-modeled large-eddy simulation of high Reynolds number flow around an airfoil near stall condition,} \emph{Computers \& Fluids}, Vol.~85, 2013, pp. 105--113.

\bibitem[{Mary and Sagaut(2002)}]{mary2002large}
Mary, I., and Sagaut, P., \enquote{Large eddy simulation of flow around an airfoil near stall,} \emph{AIAA journal}, Vol.~40, No.~6, 2002, pp. 1139--1145.

\bibitem[{Chivaee(2014)}]{chivaee2014large}
Chivaee, H.~S., \emph{Large eddy simulation of turbulent flows in wind energy}, DTU Vindenergi, 2014.

\bibitem[{Witherden et~al.(2014)Witherden, Farrington, and Vincent}]{witherden2014pyfr}
Witherden, F.~D., Farrington, A.~M., and Vincent, P.~E., \enquote{PyFR: An open source framework for solving advection--diffusion type problems on streaming architectures using the flux reconstruction approach,} \emph{Computer Physics Communications}, Vol. 185, No.~11, 2014, pp. 3028--3040.

\bibitem[{Geuzaine and Remacle(2020)}]{gmsh}
Geuzaine, C., and Remacle, J.-F., \enquote{Gmsh,} , 2020.
\newblock \urlprefix\url{http://http://gmsh.info/}.

\bibitem[{Choi and Moin(2012)}]{choi2012grid}
Choi, H., and Moin, P., \enquote{Grid-point requirements for large eddy simulation: Chapman’s estimates revisited,} \emph{Physics of fluids}, Vol.~24, No.~1, 2012, p. 011702.

\bibitem[{Said and Dickey(1984)}]{said1984testing}
Said, S.~E., and Dickey, D.~A., \enquote{Testing for unit roots in autoregressive-moving average models of unknown order,} \emph{Biometrika}, Vol.~71, No.~3, 1984, pp. 599--607.

\bibitem[{Frere et~al.(2015)Frere, Hillewaert, Chivaee, Mikkelsen, and Chatelain}]{frere2015cross}
Frere, A., Hillewaert, K., Chivaee, H.~S., Mikkelsen, R.~F., and Chatelain, P., \enquote{Cross-validation of numerical and experimental studies of transitional airfoil performance,} \emph{33rd Wind Energy Symposium}, 2015, p. 0499.

\end{thebibliography}
